\documentclass[aps,prl,reprint,groupedaddress]{revtex4-1}

\usepackage{graphicx,amsfonts,amsmath,amssymb}

\usepackage{color}

\begin{document}
\catcode`\ä = \active \catcode`\ö = \active \catcode`\ü = \active
\catcode`\Ä = \active \catcode`\Ö = \active \catcode`\Ü = \active
\catcode`\ß = \active \catcode`\é = \active \catcode`\è = \active
\catcode`\ë = \active \catcode`\ô = \active \catcode`\ê = \active
\catcode`\ø = \active \catcode`\ò = \active \catcode`\í = \active
\defä{\"a} \defö{\"o} \defü{\"u} \defÄ{\"A} \defÖ{\"O} \defÜ{\"U} \defß{\ss} \defé{\'{e}}
\defè{\`{e}} \defë{\"{e}} \defô{\^{o}} \defê{\^{e}} \defø{\o} \defò{\`{o}} \defí{\'{i}}


\title{Strongly Interacting Isotopic Bose-Fermi Mixture Immersed in a Fermi Sea}


\author{Cheng-Hsun Wu, Ibon Santiago, Jee Woo Park, Peyman Ahmadi, and Martin W. Zwierlein}
\affiliation{Department of Physics, MIT-Harvard Center for Ultracold Atoms,
and Research Laboratory of Electronics, Massachusetts Institute of Technology,
Cambridge, Massachusetts 02139, USA}

\date{\today}

\begin{abstract}
We have created a triply quantum degenerate mixture of bosonic $^{41}$K and two fermionic species $^{40}$K and $^6$Li. The boson is shown to be an efficient coolant for the two fermions, spurring hopes for the observation of fermionic superfluids with imbalanced masses. We observe multiple heteronuclear Feshbach resonances, in particular a wide $s$-wave resonance for the combination $^{41}$K-$^{40}$K, opening up studies of strongly interacting {\it isotopic} Bose-Fermi mixtures. For large imbalance, we enter the polaronic regime of dressed impurities immersed in a bosonic or fermionic bath.
\end{abstract}

\maketitle
Strongly interacting quantum mixtures of ultracold atoms provide an extremely rich platform for the study of many-body physics. They offer control over macroscopic quantum phenomena in and out of equilibrium, enabling a direct quantitative comparison to theoretical models~\cite{bloc08review}. Two-state mixtures of fermionic atoms near Feshbach resonances allow the creation of fermionic superfluids in the crossover between Bose-Einstein condensation and Bardeen-Cooper-Schrieffer (BCS) superfluidity~\cite{ingu08varenna,gior08review}. Combining different atomic species gives access to Bose-Bose~\cite{bloc01symp,modu01}, Bose-Fermi~\cite{schr01,trus01,hadz02,roat02,silb05,chin10feshbach}, and Fermi-Fermi mixtures~\cite{tagl08fermifermi,spie09fermifermi,tiec10fermifermi} that each connect to many different areas in condensed matter, high energy or nuclear physics. Bose-Fermi mixtures may provide insight into, for example, boson-mediated Cooper pairing~\cite{heis00,bijl00phonon}, QCD matter~\cite{maed09qcd}, and into theoretical models of High-$T_c$ superconductivity~\cite{lee06hightc}.
A mixture of two different fermions might allow access to a superfluid of unlike fermions. In contrast to superconductors or neutron stars, superfluid pairing will occur between particles that are not related via time-reversal symmetry.
Very recently, Fermi-Fermi mixtures of unlike fermionic species have been brought into the strongly interacting regime~\cite{tren11fermifermihydro}, offering prospects to observe universal physics in imbalanced mixtures, such as universal transport~\cite{somm11spin}.

An important class of many-body problems involves the interaction of impurities with a Fermi sea or a bosonic bath, dressing them into quasi-particles known as polarons.
For the Fermi polaron, an impurity interacting with a fermionic environment, the resulting energy shift has been experimentally measured~\cite{schi09polaron} and calculated~\cite{chev06polaron,comb07polaron,prok08polaron}. Due to the fermionic nature of the environment, the effective mass is only weakly enhanced~\cite{prok08polaron,shin08eos,nasc09imbal} even for resonant interactions. However, if the impurity swims in a bosonic bath, there is no limit in the number of bosons that interact at close distance with the impurity, and the mass-enhancement can be enormous~\cite{feyn55polaron}.

\begin{figure}
\begin{center}
\includegraphics[width=3.2in]{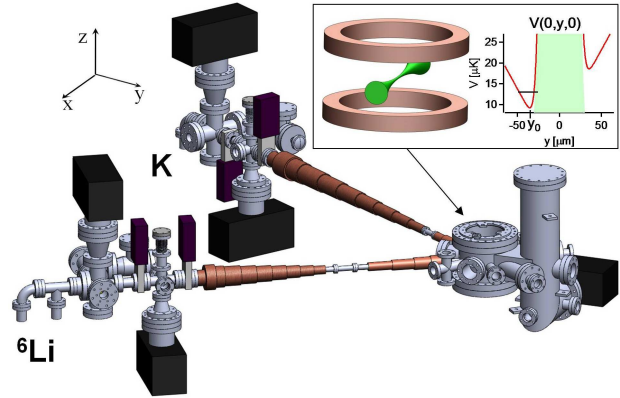}
\end{center}
\caption {Schematic of the experimental setup. Two Zeeman slowers yield optimized atom flux for $^6$Li and K, allowing a no-compromise approach to simultaneous magneto-optical trapping of $^{41}$K, $^{40}$K, and $^6$Li in the main chamber. All species are subsequently loaded into an optically plugged magnetic trap (inset). rf-evaporation of $^{41}$K sympathetically cools the fermionic species.
(inset) The trapping potential, essentially identical for all species, is sketched along the horizontal $y$-axis perpendicular to the plug beam.}\label{setup}
\end{figure}

In this work we present a rather ideal system to study strongly interacting quantum mixtures of different atomic species, a heavy, isotopic Bose-Fermi mixture of $^{40}$K-$^{41}$K with widely tunable interactions coexisting with a light Fermi sea of $^6$Li. $^{41}$K is used to sympathetically cool both $^6$Li and $^{40}$K, leading to a triply degenerate quantum mixture. Apart from a strong p-wave Feshbach resonance, we find a wide s-wave Feshbach resonance between the potassium isotopes. At our lowest temperatures at the Feshbach resonance, the mixture should be in a regime where both Bose and Fermi polarons exist. Finally, the mass-imbalanced Bose-Fermi mixture $^{6}$Li-$^{41}$K also allows for tunable interactions at several Feshbach resonances.

Predating our work, Feshbach resonances in Bose-Fermi mixtures were found in $^{23}$Na-$^6$Li, $^{87}$Rb-$^{40}$K, and in Rb-$^6$Li~\cite{chin10feshbach,deh10giant}. However, these systems are plagued by typically unequal trapping potentials and the large mass difference between unlike atoms, causing gravitational sag that has to be compensated. An atom-molecule mixture of $^6$Li-$^6$Li$_2$ allowed access to a part of the phase diagram of strongly interacting bosons and fermions~\cite{shin08bosefermi}. However, for too strong an interaction the composite nature of the bosonic molecules becomes apparent. With $^{40}$K-$^{41}$K, we have a Bose-Fermi mixture at our disposal with identical external potentials and essentially equal mass for bosons and fermions, so that the only relevant difference lies in quantum statistics.

Cooling of spin-polarized fermions to degeneracy is hindered by insufficient thermalization due to the lack of ``head-on" collisions in the s-wave regime at low temperatures, rendering Fermi gas experiments more complex than their bosonic counterparts. Common solutions include direct cooling of two hyperfine spin populations, or sympathetic cooling using another atom as a coolant~\cite{ingu08varenna}. The latter method has the advantage of mostly conserving the fermionic species in the cooling process~\cite{hadz03big_fermi}. Adverse spin-changing collisions between the coolant and the fermionic atoms can still reduce the fermion number~\cite{silb05,tagl08fermifermi}. Here we employ $^{41}$K for sympathetic cooling and find it to be an efficient coolant for both $^{40}$K and $^{6}$Li, with negligible loss of fermions.

The experimental setup, shown in Fig.~\ref{setup}, consists of two independent Zeeman slowers for lithium and potassium, allowing us to simultaneously load large samples of each of the three atomic species directly
into a UHV chamber. Although the natural abundance of $^{40}$K is only 0.01\%, the Zeeman slower with a typical flux of $10^{11}$ atoms/s for abundant species still yields $5\times 10^7$ $^{40}$K atoms loaded within two seconds into the magneto-optical trap. The same slower yields $3 \times 10^9$ $^{41}$K atoms (natural abundance 7$\%$) loaded in 2 s. We can trap $10^9$ $^6$Li atoms within 1 s.

To increase the initial atom density, a 40 ms compressed MOT phase and a 6 ms optical molasses stage compresses and cools each gas before loading into the magnetic trap. For $^{41}$K, we follow closely the procedure laid out in~\cite{kish09BEC41K}. $^{40}$K and $^6$Li require less care, as we deliberately co-trap only a few $10^5$ fermionic atoms with the coolant. The maximum number of fermions that can be brought into degeneracy by a given bosonic coolant is roughly given by the number of degenerate bosons the apparatus can provide. For $^{41}$K, this limits the fermion number to about $2 \times 10^5$, while for $^{23}$Na, the number can be as large as $7 \times 10^7$~\cite{hadz03big_fermi}.

\begin{figure}[h]
\begin{center}
\includegraphics[width=3.2in]{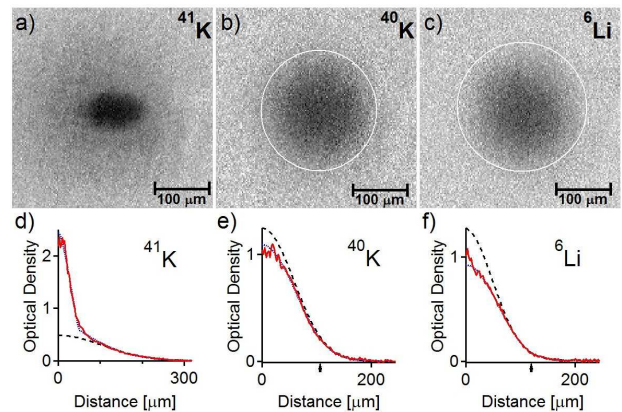}
\end{center}
\caption{a)-c): Absorption images of triply degenerate quantum gases of
$^{41}$K, $^{40}$K and $^{6}$Li, imaged after 8.12 ms, 4.06 ms and 1 ms time of flight from the magnetic trap, respectively.
The final rf-knife frequency was 500 kHz above the 254.0 MHz hyperfine transition of $^{41}$K. The white circles indicate the Fermi radius in time of flight $t$, $R_F = \sqrt{2E_F/m}\; t$.
d)-f): Azimuthally averaged column density. Solid dots: gaussian fit to the wings of the column density. Solid black and blue lines are gaussian and Fermi-Dirac fits to the entire profile. The deviation of the gaussian fit from the data is more pronounced for the more deeply degenerate $^6$Li at $T/T_F = 0.16$ than for $^{40}$K at $T/T_F = 0.51$. The arrows indicate the Fermi radii.
The atom numbers for $^{6}$Li, $^{41}$K and $^{40}$K are $1.6\times10^5$, $1.1\times10^5$, and $2.0\times10^5$, respectively.}\label{degeneracy}
\end{figure}

After the molasses stage, atoms are prepared in the stretched hyperfine states of  $|F,m_F\rangle = |2,2\rangle$ for $^{41}$K, $|9/2,9/2\rangle$ for $^{40}$K, and $|3/2,3/2\rangle$ for $^6$Li via optical pumping.
Evaporative cooling of $^{41}$K is performed in a quadrupole magnetic trap with a $B_z'= 220$ G/cm ($B_\perp'=110$ G/cm) magnetic field gradient along the vertical (horizontal) direction. To avoid Majorana spin flips, the magnetic field zero is ``plugged" by a repulsive laser beam (power 15W, wavelength 532 nm) focused to a waist of 20 $\mu$m~\cite{davi95bec}.
Unwanted hyperfine states from imperfect optical pumping are removed by reducing $B_z'$ for 200 ms to 15 G/cm, only supporting stretched states sufficiently against gravity. Without this cleaning procedure, spin-changing collisions would strongly reduce the atom number during evaporation.
Evaporation is performed on $^{41}$K by driving $|2,2\rangle \rightarrow |1,1\rangle$ rf-transitions above the hyperfine transition of 254.0 MHz. For the last 2 s of evaporation, the trap is decompressed to $B_z'= 110$ G/cm to suppress three-body losses.
A well-centered plugged trap allows for two trap minima on each side of the plug laser (see Fig.~\ref{setup}). To obtain only a single trap minimum, in the final 2 s of evaporation a horizontal bias field is applied in the $y$-direction, perpendicular to the plug beam, thus displacing the center of the magnetic trap by 10$\mu$m.
The resulting trapping potential, shown in the inset of Fig.~\ref{setup}, is approximately harmonic for atoms at energies of $\lesssim 2 \mu$K. The effect of anharmonicities is strongest along the $y$-direction, and most important for the light fermion $^{6}$Li at a typical Fermi energy of $E_F = k_B \cdot 5 \,\mu K$ ($^{40}$K only has $E_F \approx k_B \cdot 1.5 \,\mu K$).

Even for anharmonic traps, long time of flight expansion reveals the momentum distribution of the gas~\cite{kett08varenna}. Time of flight images of triply degenerate quantum mixtures are shown in Fig.~\ref{degeneracy}. Condensation of $^{41}$K is observed at $T_c = 1.2 \;\mu$K with $3 \times 10^5$ atoms. In the harmonic approximation, this translates into a geometric mean of the trapping frequencies of $\bar{\omega}_{^{41}\rm K} =2\pi \cdot 380$ Hz. Observing a $^{41}$K Bose condensate in thermal contact with a cloud of $^{40}$K and $^{6}$Li fermions each of roughly the same atom number already implies degeneracy of the fermionic species. If $T = T_{c,^{41}\rm K}$, then $T/T_{F,^{40}\rm K} = \frac{\bar{\omega}_{^{41}\rm K}}{\bar{\omega}_{^{40}\rm K}} \frac{1}{(6 \zeta(3))^{\frac{1}{3}}} \approx 0.51$ and analogously $T/T_{F,^{6}\rm Li} = 0.2$. Taking into account anharmonicities along the y-direction for $10^5$ $^{6}$Li atoms gives a small correction to the Fermi energy of -3.5\%.
Consistent with this expectation, Thomas-Fermi fits to the time of flight distributions in Fig.~\ref{degeneracy} reveal $T/T_{F,^6\rm Li} = 0.16$ ($N_{^6\rm Li} = 2.0 \cdot 10^5$) and $T/T_{F,^{40}\rm K} = 0.51$ ($N_{^{40}\rm K} = 1.1 \cdot 10^4$), while $T/T_{C,^{41}\rm K} = 0.9$. Evaporating further to obtain essentially pure condensates, we achieve $T/T_{F,^6\rm Li} = 0.10$ for $^6$Li and $T/T_{F,^{40}\rm K} = 0.35$ for $^{40}$K.

\begin{figure}
\includegraphics[width=3.4in]{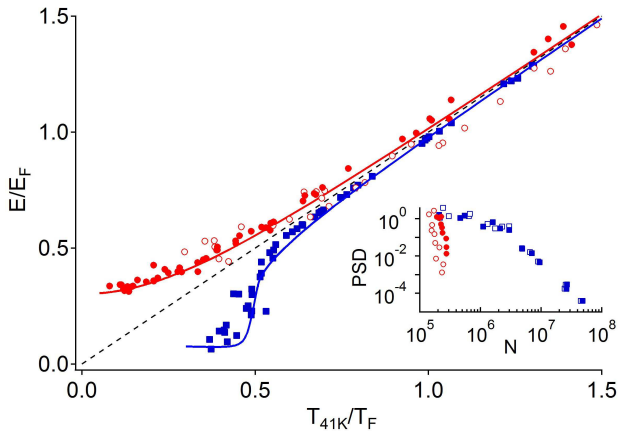}
\begin{center}
\end{center}
\caption {Observation of Pauli pressure and Bose condensation in a triply degenerate quantum mixture. Shown is the normalized release energy $E/E_F$ of each cloud versus the normalized temperature $T/T_F$. Bose condensation of $^{41}$K occurs at $T_c/T_F = 0.52$, causing a sudden reduction in release energy below $T_c$. For fermions, in contrast, the release energy saturates due to Pauli pressure. Solid circles: $^6$Li, open circles: $^{40}$K, solid squares: $^{41}$K. Solid lines: theory for an interacting Bose gas and a non-interacting Fermi gas.  Dashed line: Boltzman gas. The inset shows the evolution of the phase space density (PSD) with atom number (N) during evaporation of $^{41}$K. Open squares: Evaporation of $^{41}$K without $^6$Li and $^{40}$K.} \label{f:cooling}
\end{figure}

We directly observe Pauli pressure and Bose condensation in the triply degenerate quantum mixture. For this, we determine the $1/e$ width $R$ of a gaussian fitted to the fermionic and bosonic distributions, and compare the release energy $E \equiv \frac{1}{2}m R^2/t^2$ measured after time of flight $t$ to the Fermi energy, defined for each species as $E_F = k_B T_F = \hbar \bar{\omega} (6 N)^{1/3}$. In Fig.~\ref{f:cooling} we show $E/E_F$ as a function of the reduced temperature $T/T_F$. Thermometry is provided by fitting Bose functions to the wings of the $^{41}$K distribution. At high temperatures, $E/k_B$ simply equals the temperature of each gas. At low temperatures, the release energy of a trapped Fermi gas saturates due to Pauli pressure~\cite{trus01,schr01}, while for a Bose cloud $E$ is suddenly reduced as a condensate forms~\cite{davi95bec}.

The inset in Fig.~\ref{f:cooling} shows the phase space density (PSD) of each atom cloud versus atom number $N$ during sympathetic cooling. The efficiency of evaporation is measured by $\Gamma \equiv -{\rm d} \ln({\rm PSD})/ {\rm d} \ln(N)$. Thanks to the small fermion number, the evaporation efficiency for $^{41}$K is similar with and without load, $\Gamma \approx 3$~\cite{kish09BEC41K}. The near-vertical slope of PSD vs N for the fermionic species demonstrates efficient sympathetic cooling by $^{41}$K with $\Gamma$ = 12 (15) for $^6$Li ($^{40}$K).

We now turn to the creation of strongly interacting quantum mixtures. For this, atoms are loaded after evaporation into an optical dipole trap formed by two crossed laser beams of wavelength 1064 nm, each focussed to a waist of 100 $\mu$m at 7 W of power. For the study of $^{6}$Li-$^{41}$K Feshbach resonances, atoms are transferred into the hyperfine ground state via a Landau-Zener sweep of the bias magnetic field in the presence of 261.3 MHz and 234.2 MHz rf-radiation. For $^{40}$K-$^{41}$K, only $^{41}$K is transferred into the ground state, which is stable against spin-changing collisions due to the inverted hyperfine structure of $^{40}$K and its large nuclear spin. Feshbach resonances are detected via atom loss from three body collisions, after a fixed wait time, as a function of magnetic field. A list of observed resonances is given in Table 1.

\begin{table}
\begin{ruledtabular}
\begin{tabular}{ccccc}
Mixture & $B_0$ [G] & $\Delta B_{\rm exp}$ [G] &  Resonance type\\
\hline $^6$Li $\mid1/2,1/2\rangle$$^{41}$K$\mid1,1\rangle$ & 31.9&0.2& s-wave~\cite{hann10private}\\
\hline $^6$Li $\mid1/2,1/2\rangle$$^{41}$K$\mid1,1\rangle$ & 335.8&1.1& s-wave~\cite{hann10private}\\
\hline $^{40}$K $\mid9/2,9/2\rangle$$^{41}$K$\mid1,1\rangle$ & 472.6&0.2& s-wave~\cite{simo10private}\\
\hline $^{40}$K $\mid9/2,9/2\rangle$$^{41}$K$\mid1,1\rangle$ & 432.9&2.5 & p-wave~\cite{simo10private}\\
\hline $^{40}$K $\mid9/2,9/2\rangle$$^{41}$K$\mid1,1\rangle$ & 542.7&12& s-wave~\cite{simo10private}\\
\end{tabular}
\end{ruledtabular}
\caption{\label{tab:table3} Observed interspecies Feshbach resonances between
$^6$Li-$^{41}$K and $^{40}$K-$^{41}$K atoms. The width of the resonance, $\Delta B_{\rm exp}$, is determined by a phenomenological gaussian fit
to the observed loss feature (see e.g. Fig.\thinspace\ref{f:feshbach}). For the p-wave resonance, the width was measured at $T=8\;\mu$K.}
\end{table}

\begin{figure}[h]
\begin{center}
\includegraphics[width=3.2in]{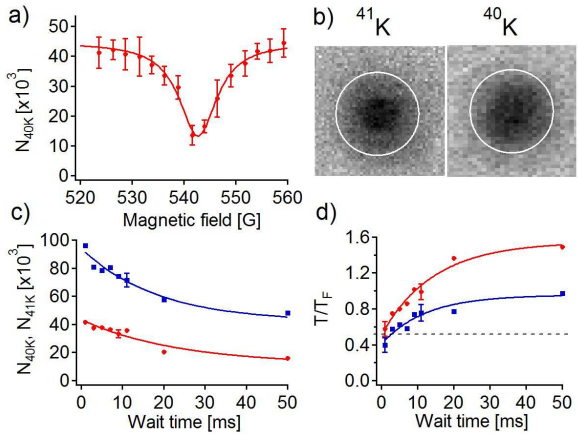}
\end{center}
\caption {Observation of a wide Feshbach resonance in the isotopic Bose-Fermi mixture of $^{41}$K-$^{40}$K. a) The atom loss feature versus magnetic field is centered at $B_0= 542.7\pm0.5$ G. b) Absorption images of the Bose and Fermi clouds after time of flight. The $^{40}$K image was scaled by the ratio of expansion factors of the Bose and Fermi cloud, the images thus approximately illustrate the in-trap density distribution. The white rim indicates the Fermi radius. c) and d) Atom number and reduced temperature $T/T_F$ versus wait time at the Feshbach resonance. Circles: $^{40}$K, squares: $^{41}$K. Dashed line: BEC threshold $T_c/T_F = 0.52$.
} \label{f:feshbach}
\end{figure}

We observe a wide Feshbach resonance in collisions of $^{40}$K in state $|9/2,9/2\rangle$ with $^{41}$K in state $|1,1\rangle$ at 543 G (Fig.~\ref{f:feshbach}a).
This resonance is theoretically predicted~\cite{simo10private} to occur at $B_0 = 541.5$ G with a width of $\Delta B=52$ G, defined via the scattering length $a = a_{\rm bg}(1+\Delta B/(B-B_0))$, where $a_{\rm bg}=65 a_0$ is the background scattering length in the vicinity of the resonance.
This isotopic Bose-Fermi mixture with essentially no gravitational sag and wide tunability of its interaction strength is very promising for controlled many-body experiments, where the only relevant difference between the two atoms is that of quantum statistics.
Fig.~\ref{f:feshbach}b) shows the immersion of a Bose-Einstein condensate of $^{41}$K into a Fermi sea of $^{40}$K with resonant interactions. The condensate survives for about 5 ms, and the remaining thermal atoms decay with a 1/e lifetime of 25 ms at initial densities $1\; (3) \times 10^{12} \;\rm cm^{-3}$ for $^{40}$K ($^{41}$K).
Our initial temperatures might be low enough, and the condensate lifetime long enough so that polarons form. At the rim of the condensate, where bosons are the minority, bosons are dressed into Fermi polarons, possibly yielding a Fermi polaron condensate~\cite{yu11bosefermi}. The formation time of such a dressed quasi-particle state should be on the order of $\hbar/E_B \sim 1\,\rm ms$, where $E_B = 0.6 E_{F,^{40}K}$ is the polaron energy~\cite{schi09polaron}.
In the center of the gas, where fermions are the minority, the gas might be in the regime where fermions are dressed by the Bose condensate.
It will be intriguing to perform local rf spectroscopy on this unconventional state of polaronic matter and to demonstrate dressing of fermionic and bosonic impurities~\cite{schi09polaron}.

In conclusion we have observed triply degenerate quantum gases of $^{41}$K, $^{40}$K and $^6$Li, through sympathetic cooling of the fermionic species by the boson $^{41}$K. In the Bose-Fermi mixtures of $^6$Li-$^{41}$K and $^{41}$K-$^{40}$K, five interspecies Feshbach resonances are detected, with s- and p-wave character. The isotopic potassium gas could become a pristine model system for strongly interacting Bose-Fermi mixtures, for example for the study of polarons~\cite{schi09polaron,nasc09imbal}, observation of polaron condensation, and universal transport of mixtures with unlike statistics~\cite{somm11spin}. The doubly degenerate $^{40}$K-$^6$Li Fermi-Fermi mixture holds promise for the observation of fermionic superfluidity and Cooper pairing between unlike fermions.
Imposing species-dependent optical potentials on mixtures will allow the study of systems with mixed dimensionality~\cite{nish08mixed} and impurity physics such as Anderson localization~\cite{gavi05} and the interaction of localized impurities with fermionic superfluids~\cite{vern11bound}.

We would like to acknowledge Andrea Simoni, Tom Hanna and Eite Tiesinga for theoretical determinations of Feshbach resonances. We also thank Sara Campbell, Christoph Clausen, Caroline Figgatt, Kevin Fisher, Vinay Ramasesh and Jacob Sharpe for experimental assistance.
This work was supported by NSF, AFOSR-MURI, ARO-MURI, ARO with funding from the DARPA OLE program, the David and Lucille Packard Foundation and the Alfred P. Sloan Foundation.

\end{document}